\documentclass[aps,prl,reprint,preprintnumbers]{revtex4-2} 
\usepackage{hyperref}
\usepackage{amsmath,latexsym}
\usepackage{graphicx}
\usepackage{siunitx}
\usepackage{textcomp}
\usepackage{wasysym}
\usepackage{amssymb}
\usepackage{MnSymbol}
\usepackage{oplotsymbl}
\usepackage{physics}
\usepackage{array}
\usepackage{tikz}

\DeclareSIUnit{\molar}{M}
\begin{document}
\title{Yield Precursor in Primary Creep of Colloidal Gels}
\date{\today}
\author{Jae Hyung Cho}
\affiliation{Department of Mechanical Engineering, Massachusetts Institute of Technology, Cambridge, Massachusetts 02139, USA}
\author{Irmgard Bischofberger}
\affiliation{Department of Mechanical Engineering, Massachusetts Institute of Technology, Cambridge, Massachusetts 02139, USA}

\begin{abstract}
Predicting the time-dependent yielding of colloidal gels under constant stress enables control of their mechanical stability and transport. Using rotational rheometry, we show that the shear rate of colloidal gels during an early stage of deformation known as primary creep can forecast an eventual yielding. Irrespective of whether the gel strain-softens or strain-hardens, the shear rate before failure exhibits a characteristic power-law decrease as a function of time, distinct from the linear viscoelastic response. We model this early-stage behavior as a series of uncorrelated local plastic events that are thermally activated, which illuminates the exponential dependence of the yield time on the applied stress. This precursor to yield in the macroscopic shear rate provides a convenient tool to identify the fate of a gel well in advance of the actual yielding.
\end{abstract}

\maketitle

\footnotetext[1]{See Supplemental Material for the experimental creep data in full ranges of the strain and the shear rate, the yield precursor at different volume fractions, and details of the model.}

Gels formed by networks of aggregated particles serve as versatile engineering materials owing to their yielding behavior. They resist deformations like solids under small stresses, but flow like liquids under stresses above a certain threshold, termed the yield stress \cite{Balmforth2014,Bonn2017}. Such a solid-to-fluid transition can occur immediately upon the application of a sufficiently large stress, or can be delayed by as long as several hours when the applied stress is just above the yield stress \cite{Rueb1997,Gopalakrishnan2007,Gibaud2010a,Laurati2011a,Sprakel2011a,Lindstrom2012,Brenner2013,Grenard2014,Leocmach2014,Calzolari2017,Aime2018c,Moghimi2021,Kusuma2021}. Whether a particulate gel under a constant stress is going to yield or not thus often remains unknown for a significant period of time, which limits our ability to prevent \cite{Blijdenstein2004,Liang2014} or harness \cite{Chang1999,Chala2018,Sahoo2018,Nelson2019,Tagliaferri2021} the time-dependent yielding. \par

Delayed yielding of particulate gels under constant stress, similar to that of numerous other systems including colloidal glasses \cite{Siebenburger2012}, microgel suspensions \cite{Uhlherr2005,Caton2008,Divoux2011}, polymer gels \cite{Karobi2016,Pommella2020} and harder materials \cite{Miguel2002,Nechad2005}, is preceded by three stages of deformation, two of which contain well-known precursors of failure \cite{Sprakel2011a,Brenner2013,Grenard2014,Aime2018c,Moghimi2021,Kusuma2021}. After the initial elastic response upon the application of the stress, the deformation continuously slows down in the first stage known as primary creep, as if the system were about to stop deforming and statically support the load without yield. Under stresses larger than the yield stress, however, secondary creep ensues during which the strain rate stays constant at a finite value. The strain rate then rapidly increases during tertiary creep, which results in the fracture of the gel network. These macroscopic signatures of approaching failure in the latter two stages concur with microscopic bursts of structural rearrangements that denote irreversible deformations \cite{Aime2018c,Cipelletti2019}. \par

In this work, we demonstrate that delayed yielding of particulate gels can be predicted by the temporal change in the rate of deformation already during \textit{primary} creep. Using rotational rheometry, we show that gels composed of attractive colloidal particles exhibit a characteristic power-law decrease in the macroscopic shear rate $\dot{\gamma}$ with time $t$ prior to yielding. Distinct from the linear viscoelastic response, this power-law decay of the shear rate is observed in a strain range independent of the applied stress $\sigma_0$. We model the macroscopic behavior as a series of mesoscopic plastic events that are thermally activated and uncorrelated in space and time. Despite its simplicity, the model reproduces the rate of change in the shear rate occurring in a stress-independent strain range and enables us to infer the exponential dependence of the yielding time on the applied stress $\sigma_0$, consistent with experimental findings. Our results hence indicate that the gradual accumulation of local plastic deformations gives rise to an early yield precursor in colloidal gels. \par

\begin{figure}[b]
\setlength{\abovecaptionskip}{-20pt}
\hspace*{-0.10cm}\includegraphics[scale=0.115]{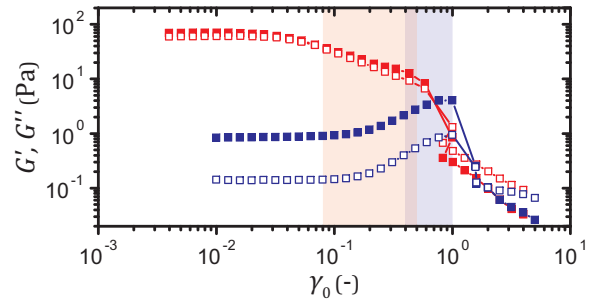}
\caption{\label{strain_swp} Storage modulus $G'$ (filled) and loss modulus $G''$ (open) as a function of the strain amplitude $\gamma_0$ at a frequency $\omega=0.63\;\si{\radian\per\s}$ for the strain-softening (red) and the strain-hardening (blue) gels at particle volume fraction $\phi=5.0\%$. The highlighted ranges of $\gamma_0$ correspond to the strain ranges in which the yield precursor emerges in creep.}
\end{figure}

\begin{figure*}[t]
\setlength{\abovecaptionskip}{-20pt}
\hspace*{-0.10cm}\includegraphics[scale=0.115]{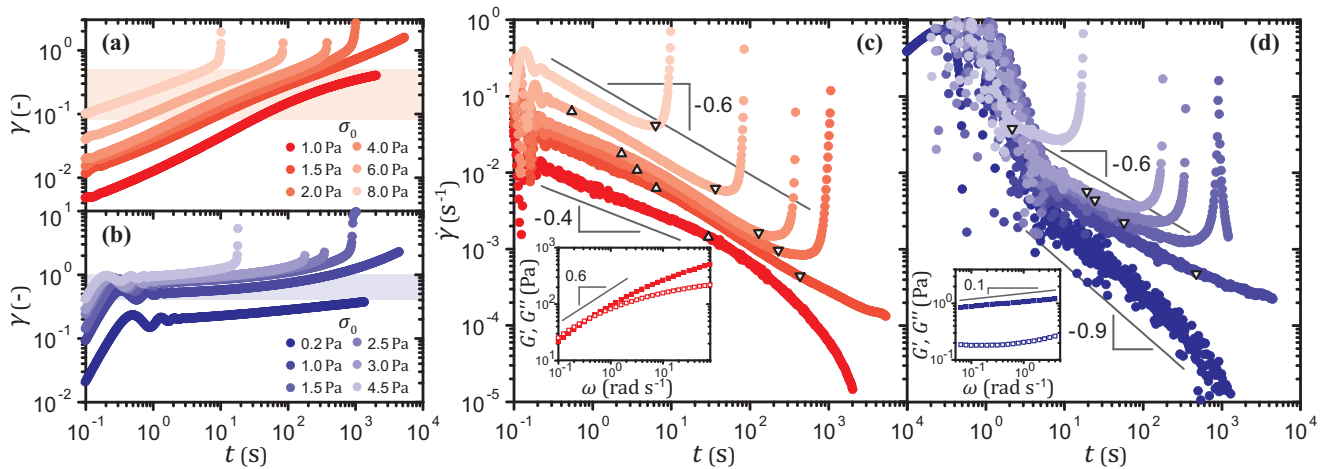}
\caption{\label{creep_exp} Strain $\gamma(t)$ of (a) the strain-softening and (b) the strain-hardening gels at $\phi=5.0\%$ during creep under different stresses $\sigma_0$ applied at time $t=0\;\si{\s}$. Shear rate $\dot{\gamma}(t)$ of (c) the strain-softening and (d) the strain-hardening gels. Triangles and inverted triangles represent $\dot{\gamma}$ corresponding to the lower and upper bounds, respectively, of the highlighted strain ranges in (a,b). Insets of (c,d): Linear viscoelastic spectra ($G'$: filled, $G''$: open).}
\end{figure*}

To show the generality of the precursor to yield in primary creep, we use two systems that exhibit markedly different yielding mechanisms under oscillatory strains, yet a nearly identical power-law decay of the shear rate under constant stresses. One system is characterized by the storage and the loss moduli, $G'$ and $G''$ respectively, that gradually decrease with the strain amplitude $\gamma_0$ after the linear regime, as displayed in Fig.~\ref{strain_swp}. For the other system, by contrast, $G'$ and $G''$ increase with $\gamma_0$, which can be ascribed to stretching of force-bearing strands \cite{Gisler1999,Colombo2014a,Bouzid2018a}, until the moduli sharply drop upon fracture, as also shown in Fig.~\ref{strain_swp}. Both the strain-softening and the strain-hardening gels are composed of polystyrene-poly(N-isopropylacrylamide) (PS-PNIPAM) core-shell particles, synthesized by an emulsion polymerization protocol \cite{Cho2021} slightly modified from those in Refs.~\cite{Dingenouts1998,Calzolari2017}. The synthesis of the strain-hardening gel particles requires an additional step that increases the thickness of the PNIPAM shells. Particles of each type are stable below the gelation temperature $T_g$, as the steric repulsion induced by the PNIPAM shells is longer-ranged than the van der Waals attraction between the PS cores. Above $T_g$, however, the particles aggregate to form gel networks, as the PNIPAM shells shrink with increasing temperature $T$, reducing the range of repulsion. The gelation temperatures $T_g$ of the strain-softening and the strain-hardening gels are $27.3\si{\celsius}$ and $25.5\si{\celsius}$ respectively \cite{Cho2021}, and we perform all experiments at $T=30\si{\celsius}$. The hydrodynamic radii $a$ of the particles measured via dynamic light scattering (BI-200SM, Brookhaven Instruments) at $T=30\si{\celsius}$ are $90.3\pm1.6\;\si{\nano\meter}$ and $116.3\pm1.8\;\si{\nano\meter}$ for the strain-softening and the strain-hardening gels respectively. We density-match the samples with a 52/48~v/v H\textsubscript{2}O/D\textsubscript{2}O mixture to prevent sedimentation and add $0.5\;\si{\molar}$ of sodium thiocyanate (NaSCN) to screen the charges of the particles. \par

We use a stress-controlled rheometer (DHR-3, TA Instruments) with a cone-plate geometry of diameter $40\;\si{\milli\meter}$. Sandpaper (grit size 600: average diameter $16\;\si{\micro\meter}$) is attached to the geometry to minimize wall slip, and the edge of the loaded sample is sealed with light mineral oil (Sigma-Aldrich) to prevent evaporation. Before each experiment, the sample is rejuvenated at a temperature $T=20\si{\celsius}<T_g$ while being presheared at a shear rate $\dot{\gamma}=500\;\si{\per\s}$ for $180\;\si{\s}$. We initiate the gelation by rapidly increasing the temperature to $T=30\si{\celsius}>T_g$ at a sample time $t_s=0\;\si{\s}$, and let the gel evolve until $t_s=2200\;\si{\s}$ before applying a constant stress $\sigma_0$, such that aging becomes negligibly slow. Both gels consist of kinetically arrested networks of uniformly sized clusters with fractal dimension $d_f=1.8\pm0.1$ \cite{Cho2020,Cho2021}, a signature of gels formed by diffusion-limited cluster aggregation \cite{Meakin1983,Weitz1984a,Weitz1984,vanDongen1985}. \par


Both types of gels yield after delays when subject to stresses slightly higher than their yield thresholds. Upon the application of a stress $\sigma_0$ at time $t=0\;\si{\s}$, the strain $\gamma(t)$ gradually increases after the initial rapid rise due to the acceleration of the instrument inertia, as shown in Fig.~\ref{creep_exp}(a,b) for a particle volume fraction $\phi=5.0\%$. For a stress $\sigma_0$ lower than the threshold, the strain $\gamma$ plateaus at late times. For $\sigma_0$ higher than the threshold, $\gamma$ rapidly increases after a delay marking macroscopic yield, once it reaches a critical value that weakly decreases with $\sigma_0$. The critical strains estimated at the elbows of the $\gamma(t)$ curves in Fig.~\ref{creep_exp}(a,b) fall within the ranges of $0.6 - 1.1$ for the strain-softening gel and $1.5 - 2.2$ for the strain-hardening gel. The yield mostly leads to fluidization of the system after which the strain $\gamma$ increases linearly with time $t$ \cite{Note1}, but strain-hardening gels that yield after delays longer than approximately $700\;\si{\s}$ can recover their stiffness such that $\gamma$ saturates to a constant, as reported for other colloidal gels that resolidify after yield \cite{Landrum2016,Moghimi2021}. For either type of gel under a stress very close to the yield stress, yielding does not occur for the duration of the experiment even when the strain increases beyond the critical range. Such a prolonged deformation, in which aging becomes no longer negligible, is outside the scope of this work. \par

\begin{figure*}[t]
\setlength{\abovecaptionskip}{-0pt}
\hspace*{-0.06cm}\includegraphics[scale=0.115]{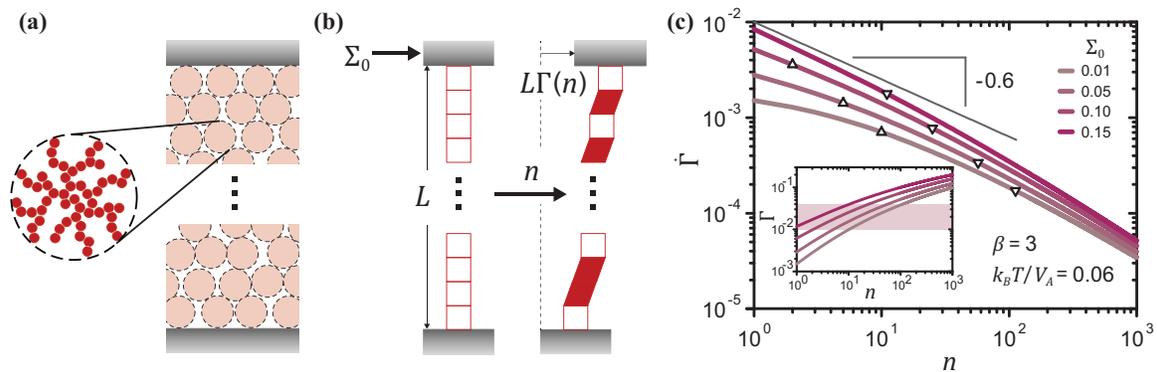}
\caption{\label{creep_model} (a) Schematic of the mesoscopic structure of colloidal gels that comprise networks of uniformly sized fractal clusters between two rigid plates. (b) One-dimensional representation of the gels as mesoscopic units connected in series with total length $L$. Each unit plastically deforms (filled parallelograms) with probability $p$ over a unit time ${\Delta}n=1$ once the stress $\Sigma_0$ is applied at time $n=0$, gradually increasing the macroscopic strain $\Gamma(n)$. (c) Shear rate $\dot{\Gamma}(n)$ and strain $\Gamma(n)$ (inset) under different $\Sigma_0$. Triangles and inverted triangles represent $\dot{\Gamma}$ corresponding to the lower and upper bounds, respectively, of the highlighted strain range $0.01-0.04$.}
\end{figure*}

Unlike the strain $\gamma(t)$, its time derivative, the shear rate $\dot{\gamma}(t)$, manifests the fate of the gel during the early stage of deformation. We find that delayed yielding, given sufficiently long primary creep, is preceded by a power-law decrease in the shear rate $\dot{\gamma}(t)$ with an exponent $-0.6$, as shown in Fig.~\ref{creep_exp}(c,d). For the strain-softening gels that do not yield, the shear rate obeys a power law $\dot{\gamma}\,\sim\,t^{-0.4}$ until it rapidly approaches zero, which reflects the viscoelastic spectrum in the inset of Fig.~\ref{creep_exp}(c). Indeed, a power law in $G^*(\omega)=1/\left[i\omega\hat{J}(\omega)\right]\,\sim\,\omega^{0.6}$, where $G^*(\omega)\,{\equiv}\,G'(\omega)+iG''(\omega)$ denotes the frequency $\omega$ dependent complex modulus and $\hat{J}(\omega)$ the Fourier transform of the creep compliance $J(t)\,{\equiv}\,\gamma(t)/{\sigma_0}$, dictates that $\dot{\gamma}(t)\,\sim\,t^{0.6-1}$ by the theory of linear viscoelasticity \cite{Bird1987a}. For the gels that do yield, the linear viscoelastic response lasts until the strain $\gamma(t)$ enters a stress-independent range $\gamma=0.08 - 0.5$ highlighted in Fig.~\ref{creep_exp}(a), which agrees with the range of strain amplitude $\gamma_0$ in which the gels strain-soften under oscillatory strains shown in Fig.~\ref{strain_swp}. When the strain lies within this range, the yield precursor $\dot{\gamma}\,\sim\,t^{-0.6}$ appears, followed by secondary creep. The strain-hardening gels exhibit similar behavior, as displayed in Fig.~\ref{creep_exp}(d). For those that do not yield, the shear rate rapidly decreases after the linear viscoelastic response $\dot{\gamma}(t)\,\sim\,t^{0.1-1}\,\sim\,t^{-0.9}$, while for those that do yield, the characteristic power law $\dot{\gamma}(t)\,\sim\,t^{-0.6}$ appears in the strain range $\gamma=0.4 - 1.0$, corresponding to the range of $\gamma_0$ in which the gels strain-harden. \par

The commonality of the yield precursor between the strain-softening and the strain-hardening gels in primary creep suggests a general mechanism for incipient plasticity in gels under constant stresses. Inspired by the development of elastoplastic models \cite{Bulatov1994,Baret2002,Bouttes2013,Merabia2016,Nicolas2018,Castellanos2018,Castellanos2019}, we build a minimal one-dimensional (1D) model that reproduces the temporal evolution of the shear rate $\dot{\Gamma}\,\sim\,{n}^{-0.6}$, where $\dot{\Gamma}$ and $n$ denote the dimensionless shear rate and time respectively, in a stress-independent range of strain $\Gamma$. 
In elastoplastic models, amorphous solids are described as networks of mesoscopic units that can locally yield \cite{Nicolas2018}. The microstructure of gels composed of networks of uniformly sized fractal clusters, as illustrated in Fig.~\ref{creep_model}(a), lends itself to such coarse-graining. \par

We model the gels as a series of mesoscopic units, the clusters, each of which bears the dimensionless applied stress $\Sigma_{0}$ during primary creep, as depicted in Fig.~\ref{creep_model}(b). We adopt several details of the model presented in Refs.~\cite{Castellanos2018,Castellanos2019} by sampling local yield thresholds $\Sigma_{y}$ from a Weibull distribution of the shape parameter $\beta$ and the scale parameter $\Sigma_{y}^{*}=1$, and by assuming thermal activation of local plastic events \cite{Note1}. The Weibull distribution can arise if the local threshold is set by the weakest link within each unit \cite{Alava2006,Bazant2019}.  At time $n=0$, each unit with a local threshold $\Sigma_{y}$ lower than the applied stress $\Sigma_{0}$ yields, resulting in the local strain $\Gamma_{l}=1$. Then per unit time, each unyielded mesoscopic unit yields with probability $p(\Sigma_y)=\nu\exp\left[-(\Sigma_{y}-\Sigma_{0})V_{a}/{k_{B}T}\right]$, where $\nu=1$ denotes the attempt frequency, $V_{a}$ the activation volume, and $k_B$ the Boltzmann constant. To focus on the early plastic behavior, we neglect elastic deformations of individual units and elastic couplings among different units, assuming no spatiotemporal correlation among plastic events. Sparse activation of plastic events in a viscous medium is indeed expected to cause minimal correlations among deformations of different fractal clusters in dilute colloidal gels. We let each unit yield only once at most. Microscopically, a plastic event of a mesoscopic unit can be understood as a straightening of originally tortuous force-bearing strands within a cluster \cite{Chan2012}. These assumptions enable us to express the average macroscopic shear rate $\dot{\Gamma}$ at time $n\geq1$ as
\begin{equation}
\dot{\Gamma}(n)=\int_{\Sigma_{0}}^{\infty} \,p(\Sigma_y)\,\left[1-p(\Sigma_y)\right]^{n-1}\,f_{wb}(\Sigma_y)\,d\Sigma_y, \label{model_sr}
\end{equation}
where $f_{wb}=\beta{\Sigma_{y}}^{\beta-1}\exp\left(-{\Sigma_{y}}^{\beta}\right)$ denotes the probability density function for the Weibull distribution of the local yield stress $\Sigma_{y}$ \cite{Note1}. The average total strain at time $n\geq1$ is $\Gamma(n)=\Gamma_0+\sum_{j=1}^{n} \dot{\Gamma}(j)$, where $\Gamma_0=1-\exp\left(-{\Sigma_0}^{\beta}\right)$ denotes the 
initial strain, equal to the cumulative distribution function of the Weibull distribution evaluated at $\Sigma_y=\Sigma_0$. \par

\begin{figure}[t]
\setlength{\abovecaptionskip}{-20pt}
\hspace*{-0.06cm}\includegraphics[scale=0.115]{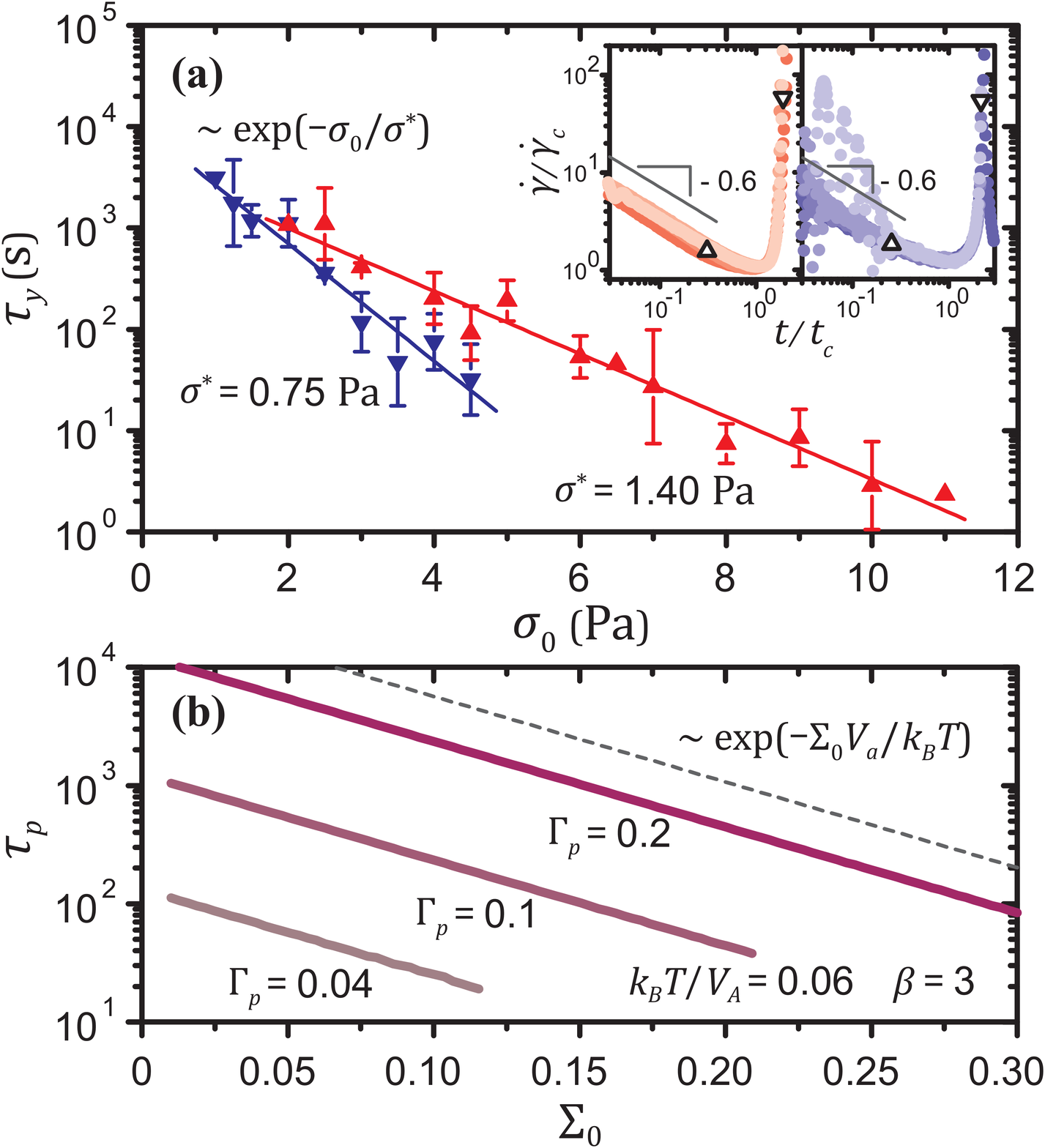}
\caption{\label{yield_time} (a) Experimental yield time $\tau_y$ as a function of the stress $\sigma_0$ for the strain-softening (triangles) and the strain-hardening (inverted triangles) gels with exponential fits at $\phi=5.0\%$. Inset: Shear rate $\dot{\gamma}(t)$ of both types of gels scaled by the minimum shear rate $\dot{\gamma}_c$ and the corresponding time $t_c$ for different $\sigma_0$. Triangles denote the ends of the yield precursor, and inverted triangles denote yield. (b) Elapsed time $\tau_p$ to reach different arbitrary total strains $\Gamma_p$ in the model, as a function of the stress $\Sigma_0$. All curves scale as $\tau_p\,\sim\,\exp\left(-{\Sigma_0}V_{a}/{k_{B}T}\right)$ (dashed line).}
\end{figure}

This model shows a near power-law decay of the shear rate within a stress-independent strain range, as experimentally found in primary creep. For the shape parameter $\beta=3$ and the thermal energy density $k_BT/{V_a}=0.06$, we observe that the shear rate follows $\dot{\Gamma}\,\sim\,{n}^{-0.6}$ within a strain range $\Gamma=0.01-0.04$ for stresses from $\Sigma_0=0.01$ to $0.15$, as shown in Fig.~\ref{creep_model}(c). Although changing the values of $\beta$ and $k_BT/{V_a}$ can alter the power-law exponent, the stress independence of the exponent within a strain range is robust, regardless of the specifics of the parameters \cite{Note1}. The monotonic decrease in $\dot{\Gamma}(n)$ is a consequence of the gradual depletion of unyielded mesoscopic units with lower yield thresholds, and hence reflects a statistical hardening effect \cite{Castellanos2019,Baret2002}. This effect explains why macroscopic plastic deformation slows down under constant stress in the experiments. Although the accumulation of yielded units should eventually induce stronger correlations among different local yielding events leading to a proliferation of plastic deformations throughout the gel network \cite{Aime2018c}, our model, which explicitly represents the onset of plasticity, continues to display statistical hardening at late times. \par

The model indicates that the thermal activation of local plastic events gives rise to the exponential dependence of the yield time on the applied stress. In the experiments, we find that both strain-softening and strain-hardening gels exhibit a yield time $\tau_y$, measured at the inflection point of the shear rate $\dot{\gamma}(t)$ for the fluidizing samples or at the local peak of $\dot{\gamma}(t)$ for the re-solidifying ones, that decreases with the stress $\sigma_0$ as $\tau_y\,\sim\,\exp\left(-\sigma_0/\sigma^*\right)$, where $\sigma^*$ denotes a characteristic stress, as shown in Fig.~\ref{yield_time}(a). Such $\sigma_0$ dependence of $\tau_y$ has been reported for various other colloidal gels \cite{Gopalakrishnan2007,Gibaud2010a,Sprakel2011a,Lindstrom2012,Brenner2013,Grenard2014,Kusuma2021}. Since the shear rate curves from the end of the primary creep to yield for different $\sigma_0$ can be collapsed onto a master curve for either type of gel as displayed in the inset of Fig.~\ref{yield_time}(a), the time at which the yield precursor ends also exponentially decreases with $\sigma_0$. The model indeed shows that the time elapsed $\tau_p$ to reach an arbitrary total strain $\Gamma_p$ scales as $\exp\left(-\Sigma_{0}V_{a}/k_{B}T\right)$, as shown in Fig.~\ref{yield_time}(b). By comparing the scalings between the experiments and the model, $\sigma^*=k_{B}T/V_{a}$, we estimate the activation radii $r_a\equiv\left(3V_a/4\pi\right)^{1/3}$ to be $89.3\pm1.8\,\si{\nano\meter}$ and $110.1\pm3.8\,\si{\nano\meter}$, surprisingly close to the particle radii $a=90.3\pm1.6\,\si{\nano\meter}$ and $116.3\pm1.8\,\si{\nano\meter}$, for the strain-softening and the strain-hardening gels respectively. This agreement hints towards particle-scale rearrangements triggering the local plastic events, reminiscent of the particle-scale plasticity observed in gels under cyclic loadings \cite{vanDoorn2018a}. \par

\begin{figure}[t]
\setlength{\abovecaptionskip}{-27pt}
\hspace*{-0.25cm}\includegraphics[scale=0.122]{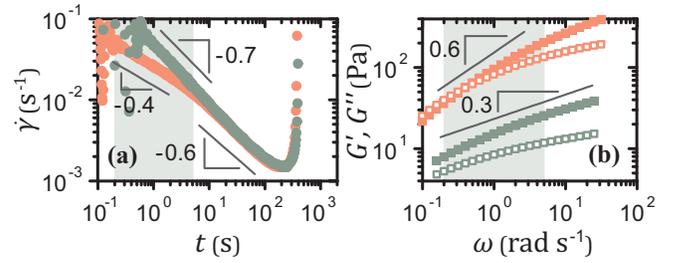}
\caption{\label{tuning} (a) Shear rate $\dot{\gamma}(t)$ of the strain-softening gel (red, $\sigma_0=4\;\si{\pascal}$) and the binary gel with a volume ratio $7:3$ of the strain-softening to the strain-hardening gel particles (green, $\sigma_0=3\;\si{\pascal}$) for $\phi=5.0\%$. (b) Linear viscoelastic spectra of the two systems. Highlighted regions indicate corresponding timescales in the two experiments ($t\sim\omega^{-1}$).}
\end{figure}

The reflection of the linear viscoelasticity and the emergent plasticity enables us to tune the time dependence of the shear rate during primary creep. We show in Fig.~\ref{tuning}(a) how $\dot{\gamma}(t)$ can be engineered by mixing the two types of particles, which alters the linear viscoelastic spectrum \cite{Cho2021}. Since the moduli of the binary gel follow $G',G''\,\sim\,\omega^{0.3}$, as displayed in Fig.~\ref{tuning}(b), the shear rate in the linear strain range decreases as $\dot{\gamma}\,\sim\,t^{-0.7}$, followed by the yield precursor $\dot{\gamma}\,\sim\,t^{-0.6}$, distinct from the responses of the original gels. \par

In summary, we demonstrate that a characteristic power-law decay of the shear rate $\dot{\gamma}(t)$ during primary creep forecasts delayed yielding of colloidal gels. A single set of the shape parameter $\beta=3$ of the local yield stress distribution and the thermal energy density $k_{B}T/V_a=0.06$ in our model is sufficient to reproduce $\dot{\gamma}(t)\,\sim\,t^{-0.6}$ that appears in our experiments for both types of PS-PNIPAM gels for different volume fractions \cite{Note1}. We note that other exponents that may better describe the behaviors of various colloidal gels can also be reproduced by tuning $\beta$ and $k_{B}T/V_a$ \cite{Note1}. These results clarify the existence of plasticity in gels during primary creep \cite{Helal2016}, and will enable more deft manipulation of yield stress fluids in diverse applications. \par

\begin{acknowledgments}
We thank Thibaut Divoux for helpful discussions. We acknowledge support from the MIT Research Support Committee and the Kwanjeong Educational Foundation, Awards No. 16AmB02M and No. 18AmB59D. 
\end{acknowledgments}

\bibliographystyle{apsrev4-1}
\bibliography{Yield_precursor_paper_Oct_21_fixed.bib}

\end{document}


\title{Yield Precursor in Primary Creep of Colloidal Gels: Supplemental Material}
\author{Jae Hyung Cho}
\affiliation{Department of Mechanical Engineering, Massachusetts Institute of Technology, Cambridge, Massachusetts 02139, USA}
\author{Irmgard Bischofberger}
\affiliation{Department of Mechanical Engineering, Massachusetts Institute of Technology, Cambridge, Massachusetts 02139, USA}

\maketitle

\subsection{1. Full ranges of the strain $\gamma(t)$ and the shear rate $\dot{\gamma}(t)$ during creep}
\begin{figure}[hb]
\setlength{\abovecaptionskip}{-20pt}
\hspace*{-0.10cm}\includegraphics[scale=0.120]{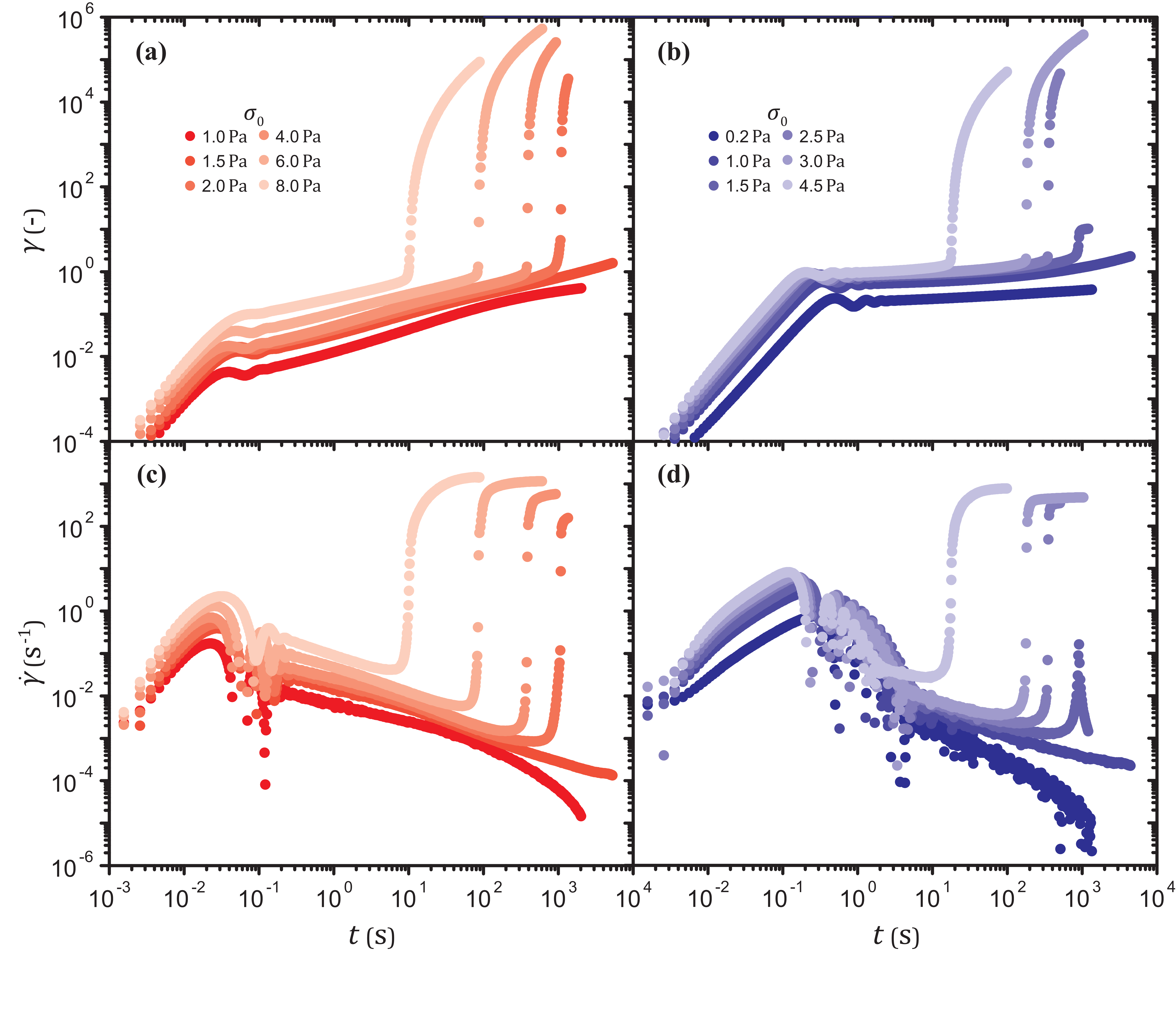}
\caption{\label{full_range} Full ranges of (a,b) the strain $\gamma(t)$ and (c,d) the shear rate $\dot{\gamma}(t)$ of (a,c) the strain-softening and (b,d) the strain-hardening gels for a particle volume fraction $\phi=5.0\%$ during creep under different stresses $\sigma_0$.}
\end{figure}

The full ranges of the strain $\gamma(t)$ and the shear rate $\dot{\gamma}(t)$ of the strain-softening and the strain-hardening gels are shown in Fig.~\ref{full_range} for a particle volume fraction $\phi=5.0\%$. The shear rate $\dot{\gamma}(t)$ after delayed yielding slowly approaches a constant value for fluidized samples. 

\newpage

\subsection{2. Yield precursor for different particle volume fractions}
\begin{figure}[hb]
\setlength{\abovecaptionskip}{-20pt}
\hspace*{-0.10cm}\includegraphics[scale=0.120]{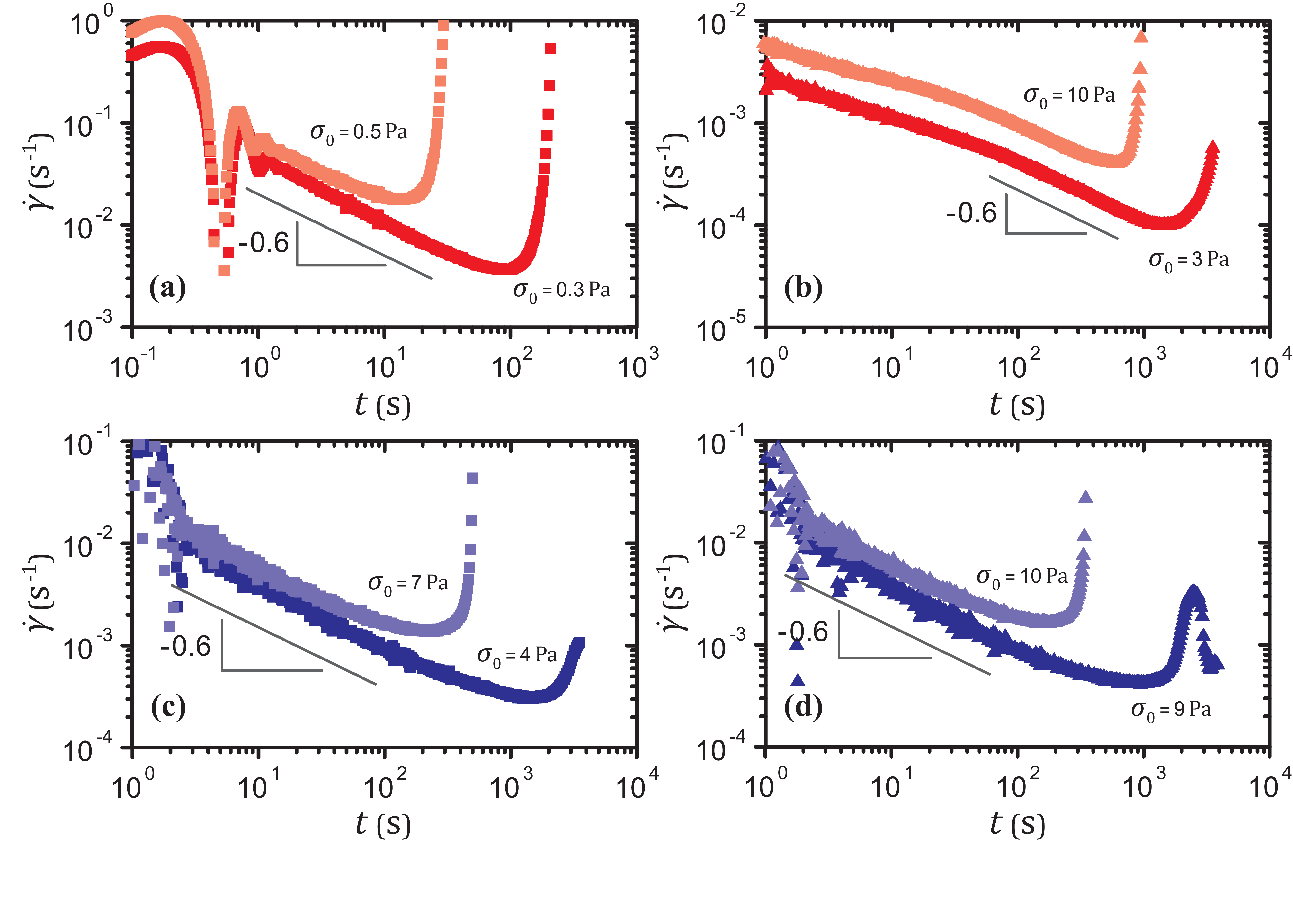}
\caption{\label{other_phi} Precursor to yield during primary creep ($\dot{\gamma}\,\sim\,t^{-0.6}$) in (a,b) the strain-softening and (c,d) the strain-hardening gels under different stresses $\sigma_0$ for particle volume fractions (a) $\phi=2.5\%$, (b,c) $7.5\%$, and (d) $10\%$.}
\end{figure}

The yield precursor during primary creep $\dot{\gamma}\,\sim\,t^{-0.6}$ found for the gels at particle volume fraction of $\phi=5.0\%$ is also observed for other volume fractions $\phi$, as displayed in Fig.~\ref{other_phi}. 

\newpage

\subsection{3. Calculation of the dimensionless shear rate $\dot{\Gamma}(n)$ in the one-dimensional model}
In our one-dimensional (1D) model for the creep deformation of colloidal gels, the macroscopic strain $\Gamma$ at dimensionless time $n$ is obtained as $\Gamma(n)=\frac{N_p(n)}{N}\Gamma_l$, where $N_p$ denotes the number of plastically deformed mesoscopic units, $N$ the total number of mesoscopic units in the system, and $\Gamma_l$ the local strain of each plastically deformed mesoscopic unit. We assume $\Gamma_l=1$ for all units such that $\Gamma$ is equal to the fraction of yielded units. Therefore, $\Gamma=1$ when all units are plastically deformed. We let each unit yield once at most. \par

\begin{figure}[hb]
\setlength{\abovecaptionskip}{-20pt}
\hspace*{-0.10cm}\includegraphics[scale=0.118]{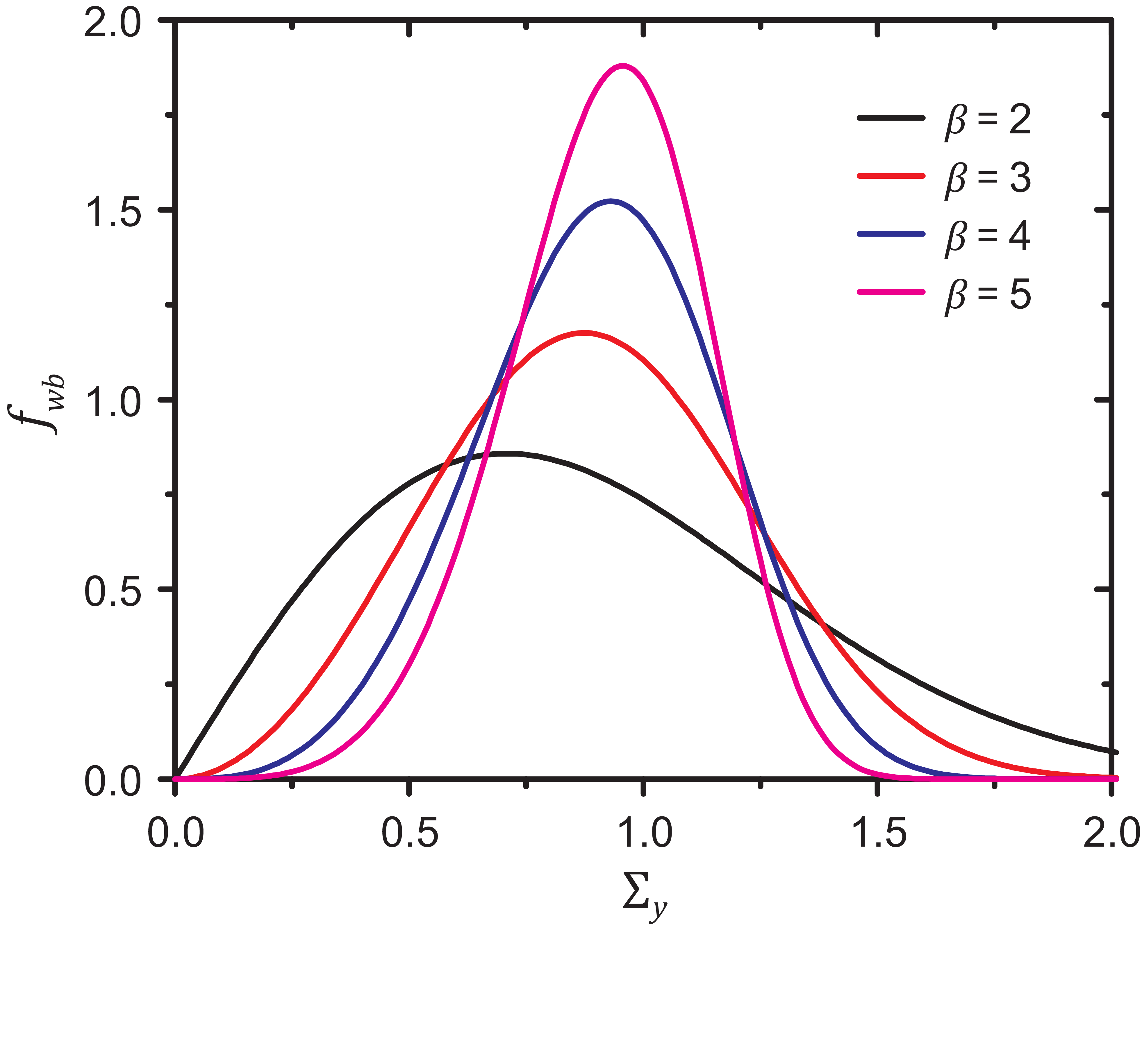}
\caption{\label{wblpdf} Probability density functions of the Weibull distribution $f_{wb}$ expressed in Eq.~\eqref{wblpdf_eq} for a scale parameter $\Sigma_y^*=1$ and shape parameters $\beta=2,3,4,5$.}
\end{figure}

Adopting the approach presented in Refs.~\cite{Castellanos2018,Castellanos2019}, the dimensionless local yield stress $\Sigma_{y}$ of each mesoscopic unit is sampled from a Weibull distribution, whose probability density function $f_{wb}$ can be expressed as 
\begin{equation}
f_{wb}(\Sigma_y)=\frac{\beta}{\Sigma_y^*}\left(\frac{\Sigma_{y}}{\Sigma_y^*}\right)^{\beta-1}\exp\left[-\left(\frac{\Sigma_y}{\Sigma_y^*}\right)^{\beta}\right], \label{wblpdf_eq}
\end{equation}
where $\beta$ denotes the shape parameter and $\Sigma_y^*$ the scale parameter. Numerical values of the local yield stress $\Sigma_y$ are insignificant, and hence we assume a scale parameter $\Sigma_y^*=1$. The greater the shape parameter $\beta$, the sharper $f_{wb}$ is around $\Sigma_y=\Sigma_y^*=1$, as shown in Fig.~\ref{wblpdf}; the local yield stresses of different units are more narrowly distributed around $\Sigma_y^*=1$ for larger $\beta$.

Upon the application of a constant dimensionless stress $\Sigma_0$ at $n=0$, every unit with a local yield stress $\Sigma_y<\Sigma_0$ plastically deforms. Given a sufficiently large number of units $N$, the resultant initial macroscopic strain $\Gamma_0$ is the same as the fraction of the units with $\Sigma_y<\Sigma_0$, or $\Gamma_0=1-\exp\left(-{\Sigma_0}^\beta\right)$, which is equal to the cumulative distribution function $F_{wb}(\Sigma_y)$ of the local yield stress evaluated at $\Sigma_y=\Sigma_0$. \par

For each unyielded unit, the probability of plastic deformation over a unit time ${\Delta}n=1$ is assumed to be
\begin{equation}
p(\Sigma_y,\Sigma_0)=\nu\exp\left[-\frac{\left(\Sigma_y-\Sigma_0\right)V_a}{k_{B}T}\right],
\end{equation}
where $\nu=1$ denotes the attempt frequency, $V_a$ the dimensionless activation volume, and $k_{B}T$ the dimensionless thermal energy. In the absence of spatiotemporal correlation, the probability of an arbitrary mesoscopic unit with $\Sigma_y>\Sigma_0$ being plastically deformed at time $n\geq1$ becomes $p^{1}\left(1-p\right)^{n-1}$. Thus, the averaged probability of local yield at $n\geq1$ for a mesoscopic unit with $\Sigma_y>\Sigma_0$ can be expressed as
\begin{equation}
P(n,\Sigma_0)=\frac{\int_{\Sigma_0}^{\infty}p(\Sigma_y,\Sigma_0)\left[1-p(\Sigma_y,\Sigma_0)\right]^{n-1}f_{wb}(\Sigma_y)\,d\Sigma_y}{\int_{\Sigma_0}^{\infty}f_{wb}(\Sigma_y)\,d\Sigma_y}=\frac{\int_{\Sigma_0}^{\infty}p(\Sigma_y,\Sigma_0)\left[1-p(\Sigma_y,\Sigma_0)\right]^{n-1}f_{wb}(\Sigma_y)\,d\Sigma_y}{1-F_{wb}(\Sigma_0)}.
\end{equation}
The average macroscopic shear rate at $n\geq1$ is equal to the number of units with $\Sigma_y>\Sigma_0$ times the macroscopic strain increment due to a single plastic event, multiplied by $P(n,\Sigma_0)$:
\begin{align}
\dot{\Gamma}(n,\Sigma_0)&=N\left[1-F_{wb}(\Sigma_0)\right]\;\times\;\frac{\Gamma_l}{N}\;{\times}\;P(n,\Sigma_0) \nonumber \\
&=\int_{\Sigma_0}^{\infty}p(\Sigma_y,\Sigma_0)\left[1-p(\Sigma_y,\Sigma_0)\right]^{n-1}f_{wb}(\Sigma_y)\,d\Sigma_y.
\end{align}
Finally, the average total strain at $n\geq1$ can be obtained from
\begin{equation}
\Gamma(n,\Sigma_0)=\Gamma_0(\Sigma_0)+\sum_{j=1}^{n}\dot{\Gamma}(j,\Sigma_0).
\end{equation}

\newpage

\subsection{4. Dependence of the shear rate $\dot{\Gamma}(n)$ on the shape parameter $\beta$ and the thermal energy density $k_{B}T/{V_a}$}

\begin{figure}[hb]
\setlength{\abovecaptionskip}{0pt}
\hspace*{-0.10cm}\includegraphics[scale=0.118]{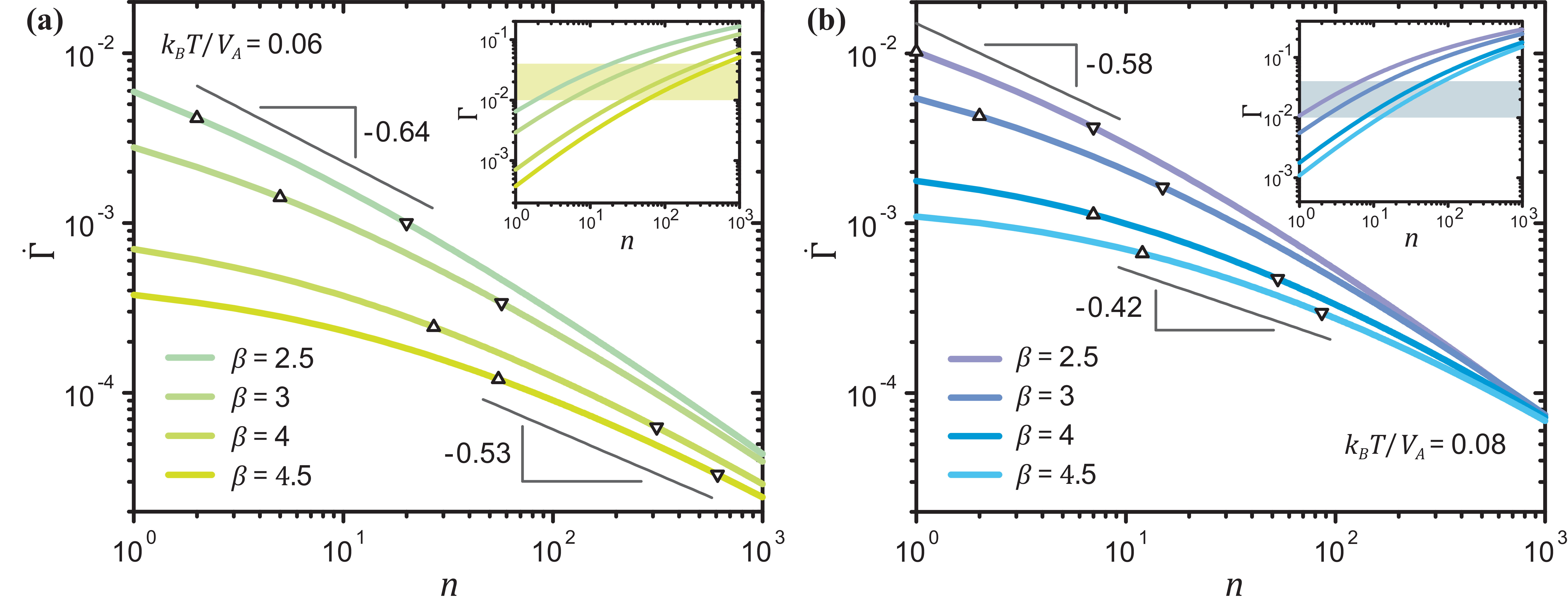}
\caption{\label{diff_beta} Dependence of the macroscopic shear rate $\dot{\Gamma}(n)$ on the shape parameter $\beta$ for a fixed stress $\Sigma_0=0.05$ and fixed thermal energy densities $k_{B}T/V_a$ of (a) $0.06$ and (b) $0.08$. Insets: Corresponding macroscopic strain $\Gamma(n)$. For each curve, the triangle and the inverted triangle represent the shear rates $\dot{\Gamma}$ at which the strain $\Gamma$ enters and exits the range $0.01-0.04$, respectively.}
\end{figure}

In the main text, the shear rate $\dot{\Gamma}$ in the model is shown to exhibit a power law $\dot{\Gamma}\,\sim\,n^{-0.6}$, independent of the applied stress $\Sigma_0$, when the strain falls within the range of $\Gamma=0.01 - 0.04$ for the shape parameter $\beta=3$ and the thermal energy density $k_{B}T/V_a=0.06$. Larger values of $\beta$ result in smaller values of the magnitude of the power-law exponent, as shown in Fig.~\ref{diff_beta}(a) for the range $\Gamma=0.01-0.04$. A larger $\beta$ corresponds to a more narrow distribution of the local yield stress around $\Sigma_y=1$, as displayed in Fig.~\ref{wblpdf}, which reduces the fraction of mesoscopic units whose local yield stresses are close to the applied stress $\Sigma_0\ll1$. The scarcity of units with lower yield thresholds renders plastic events unlikely from the outset, and hence partially obscures the statistical hardening, leading to a smaller magnitude of the power-law exponent. Such $\beta$ dependence of the exponent is more clearly observed for a higher thermal energy density $k_{B}T/V_a=0.08$, as displayed in Fig.~\ref{diff_beta}(b) for the same strain range $\Gamma=0.01-0.04$. \par

\begin{figure}[hb]
\setlength{\abovecaptionskip}{0pt}
\hspace*{-0.10cm}\includegraphics[scale=0.118]{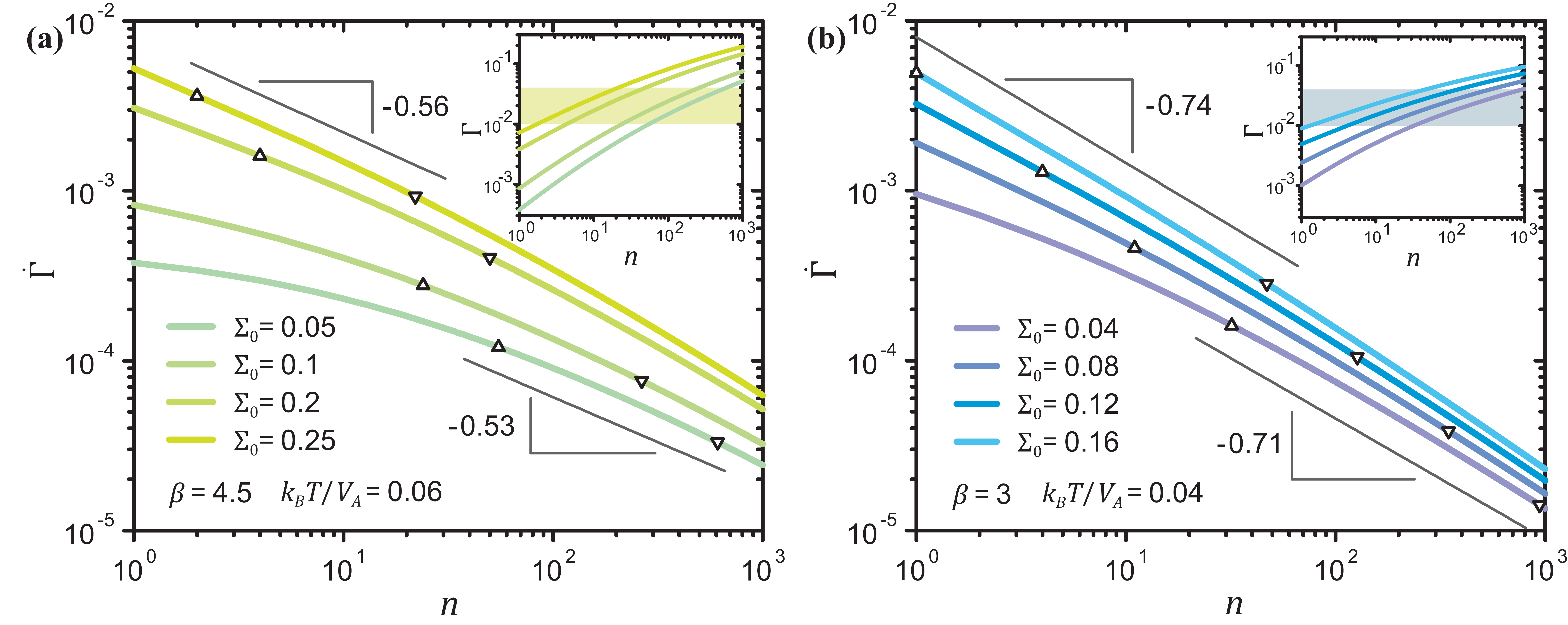}
\caption{\label{diff_sig} Dependence of the macroscopic shear rate $\dot{\Gamma}(n)$ on the stress $\Sigma_0$ for (a) $\beta=4.5$, $k_{B}T/V_a=0.06$ and (b) $\beta=3$, $k_{B}T/V_a=0.04$. Insets: Corresponding macroscopic strain $\Gamma(n)$. For each curve, the triangle and the inverted triangle represent the shear rates $\dot{\Gamma}$ at which the strain $\Gamma$ enters and exits the range $0.01-0.04$, respectively.}
\end{figure}

For any given set of $\beta$ and $k_{B}T/V_a$, however, the applied stress $\Sigma_0$ minimally alters the power-law exponent of $\dot{\Gamma}(n)$ within a fixed strain range. Although selections of the parameter values other than $\beta=3$ and $k_{B}T/V_a=0.06$ change the exponent to a value different from $-0.6$ in the strain range $\Gamma=0.01-0.04$, the exponent remains largely independent of $\Sigma_0$, as shown in Fig.~\ref{diff_sig}(a,b). Our model can therefore be used to account for exponents other than $-0.6$, which may better describe the deformation rates in other soft materials. 
 
\bibliographystyle{apsrev4-1}
\bibliography{Yield_precursor_paper_Oct_21_fixed.bib}